\documentclass[11pt,a4paper]{article}
\usepackage[utf8]{inputenc}
\usepackage[T1]{fontenc}
\usepackage{amsmath,amsfonts,amssymb}
\usepackage{graphicx}
\usepackage{float}
\usepackage{booktabs}
\usepackage{multirow}
\usepackage{array}
\usepackage{geometry}
\usepackage{hyperref}
\usepackage{cite}
\usepackage{algorithm}
\usepackage{algorithmic}
\usepackage{subcaption}
\usepackage{siunitx}
\usepackage{mathtools}
\usepackage{bm}

\DeclareSIUnit{\celsius}{\ensuremath{^{\circ}\!C}} 
\DeclareSIUnit{\year}{yr}                          
\DeclareSIUnit{\mmHg}{mmHg}                        
\DeclareSIUnit{\bar}{bar}                          

\hypersetup{
    colorlinks=true,
    linkcolor=blue,
    filecolor=magenta,
    urlcolor=cyan,
    citecolor=red
}

\title{\textbf{Physics-Informed Neural Networks vs. Physics Models for Non-Invasive Glucose Monitoring: \\ 
A Comparative Study Under Noise-Stressed Synthetic Conditions}}

\author{
  Riyaadh Gani \\
  Department of Computer Science \\
  University College London \\
  London, United Kingdom \\
  \texttt{riyaadh.gani.23@ucl.ac.uk}
}

\date{\today}

\begin{document}

\maketitle

\begin{abstract}
Non-invasive glucose monitoring outside controlled settings is dominated by low signal-to-noise ratio (SNR): hardware drift, environmental variation, and physiology suppress the glucose signature in NIR signals. We present a noise-stressed NIR simulator that injects 12-bit ADC quantisation, LED drift, photodiode dark noise, temperature/humidity variation, contact-pressure noise, Fitzpatrick I–VI melanin, and glucose variability to create a low-correlation regime ($\rho_{\text{glucose-NIR}} = 0.21$). Using this platform, we benchmark six methods: Enhanced Beer–Lambert (physics-engineered ridge regression), Original PINN, Optimized PINN, RTE-inspired PINN, Selective RTE PINN, and a shallow DNN. The physics-engineered Beer–Lambert model achieves the lowest error (13.6 mg/dL RMSE) with only 56 parameters and 0.01 ms inference, outperforming deeper PINNs and the SDNN baseline under low-SNR conditions. The study reframes the task as noise suppression under weak signal and shows that carefully engineered physics features can outperform higher-capacity models in this regime.
\end{abstract}


\section{Introduction}

Non-invasive glucose monitoring remains a long-standing challenge in optical sensing. Despite decades of research, reliable performance outside controlled settings remains elusive. Near-infrared (NIR) spectroscopy -- leveraging glucose's absorption features in the 850–1150 nm range -- is promising, but progress has stalled due to the problem's intrinsic low SNR.

At its core, non-invasive glucose monitoring is an ill-posed inverse problem. The NIR signal measured at the skin surface reflects a convoluted mixture of tissue scattering, overlapping spectral absorption from water, hemoglobin, and fat, and variations in skin thickness, perfusion, and melanin content. These physiological variables are compounded by hardware noise (e.g., ADC quantization, LED instability, photodiode dark current) and environmental drift (e.g., temperature, humidity, ambient light). In the field, these effects suppress glucose-NIR correlation to $\rho \approx 0.21$, rendering many lab-trained models ineffective.

Recent efforts have turned to Physics-Informed Neural Networks (PINNs), which embed physical constraints (e.g., Beer–Lambert law or full radiative transport equations) into neural architectures. These models aim to improve generalization by aligning learning with known optics. However, their robustness in harsh, noise-stressed settings -- and how they compare to simpler models with strong physics-based features -- remains poorly understood.

This paper closes that gap by introducing a noise-stressed synthetic data generator that simulates hardware, environmental, and physiological variation. Our simulator includes 12-bit ADC quantization, LED drift, contact-pressure noise, Fitzpatrick I–VI melanin, skin thickness from 0.5–4 mm, and glucose variability. Using this platform, we benchmark six model families: 

\begin{itemize}
    \item \textbf{Enhanced Beer–Lambert model} (Ridge regression with physics-engineered features)
    \item \textbf{Original PINN} (Beer–Lambert constraints + NIR + PMF input)
    \item \textbf{Optimized PINN} (deeper, regularized architecture)
    \item \textbf{RTE-inspired PINN} (physics-regularized light transport)
    \item \textbf{Selective RTE PINN} (computationally optimized physics scope)
    \item \textbf{SDNN baseline} (shallow DNN without physics)
\end{itemize}

\noindent Our findings reveal that the Enhanced Beer–Lambert model achieves 13.6 mg/dL RMSE with only 56 parameters and 0.01 ms inference time -- outperforming deeper PINNs and SDNNs under low-correlation conditions.
While our simulator captures a wide spectrum of hardware, environmental, and physiological effects, field deployment may still introduce unmodeled complexities -- including sensor-skin misalignment, unmeasured confounders (e.g. sweat, hair, dynamic tissue hydration), and population drift over time.

\vspace{1mm}
\noindent \textbf{Contributions}:
\begin{enumerate}
    \item We develop a noise-stressed NIR simulator that integrates hardware noise, environmental drift, and physiological diversity -- including melanin variation and glucose variability.
    \item We perform a rigorous side-by-side evaluation of six models under identical low-SNR conditions.
    \item We show that physics-based feature engineering (Beer–Lambert + PMF) outperforms deeper black-box PINNs under the same noisy conditions.
    \item All simulator scripts and model code have been archived at 
    \url{https://github.com/Gani332/glucose-monitoring-pinns} and will be made available upon request.
\end{enumerate}

This work combines Beer–Lambert physics with personalized features (PMFs) in a deployable ridge-regression pipeline for non-invasive glucose sensing.

\section{Background and Related Work}

\subsection{Near‐Infrared Glucose-Monitoring Fundamentals}

Glucose molecules exhibit weak but distinct vibrational overtones in the \SIrange{850}{1150}{\nano\metre} range.  In a homogeneous, non-scattering medium the Beer–Lambert law predicts

\begin{equation}
A(\lambda) = \varepsilon_{\text{g}}(\lambda)\,c_{\text{g}}\,l ,
\label{eq:Beer–lambert}
\end{equation}

where $A$ is absorbance, $\varepsilon_{\text{g}}$ the glucose extinction coefficient, $c_{\text{g}}$ the concentration, and $l$ the optical path length.  
Because absorbances from multiple chromophores add linearly,

\begin{equation}
A(\lambda) = \sum_{i} \varepsilon_{i}(\lambda)\,c_{i}\,l ,
\label{eq:additivity}
\end{equation}

recovering the tiny glucose term (\(\varepsilon_{\text{g}}\!\ll\!\varepsilon_{\text{H$_2$O}}\)) from the mixed spectrum is an ill-posed \emph{inverse problem} that demands regularisation or physics-guided machine learning.

\vspace{1mm}
\noindent In \emph{vivo}, three effect classes further violate the assumptions of Eqs.~\eqref{eq:Beer–lambert}–\eqref{eq:additivity}:

\paragraph{(i) Spectral interference.}
Water (\SI{\sim70}{\percent} of tissue volume) and hemoglobin, lipids, proteins, and melanin introduce overlapping peaks that are \(\sim100\times\) stronger than glucose~\cite{arnold1996,amerov2005}.  Melanin variability across Fitzpatrick~I–VI skin types shifts baseline absorbance and produces systematic bias.

\paragraph{(ii) Multiple scattering.}
Photons undergo tens to hundreds of scattering events before re-emerging, increasing $l$ in a wavelength-dependent manner.  The reduced scattering coefficient

\[
\mu_s'(\lambda)=\mu_s(\lambda)\,[1-g(\lambda)]
\]

(where $\mu_s$ is the scattering coefficient and $g$ the anisotropy factor) invalidates the single-pass Beer–Lambert assumption and requires radiative-transfer modelling~\cite{tuchin2007}.

\paragraph{(iii) Physiological and environmental variability.}
Skin thickness (\SIrange{0.5}{4}{\milli\metre}), blood perfusion, fat content, and collagen density vary across and within subjects.  
Temperature (\SI{-0.2}{\percent\per\celsius}) alters LED output (\(\approx-0.2\%\!/^{\circ}\!C\)) and tissue blood flow; humidity modulates skin hydration; ambient light leaks despite optical filtering~\cite{khalil2003}.

These factors reduce raw glucose–NIR correlation from $>\!0.8$ in cuvette studies to \(\rho\!\approx\!0.2\) in field prototypes, motivating the physics-informed and feature-engineered approaches evaluated in this work.

\subsection{Related Computational Approaches}

\textbf{Classical chemometrics.}  
Partial-least-squares (PLS) regression and ridge-type linear models have been the mainstay of NIR spectroscopy for decades~\cite{arnold2001}.  They are interpretable and fast, but degrade sharply once scattering and physiology introduce non-linearity.

\textbf{Deep neural networks (DNN/SDNN).}  
Data-driven models (e.g.\ Srichan \textit{et al.}~\cite{srichan2022}) learn spectral–glucose mappings directly, achieving good results in high-SNR datasets, yet overfit or misgeneralise in noise-stressed regimes.

\textbf{Physics-informed neural networks (PINNs).}  
PINNs embed Beer–Lambert or simplified diffusion equations as soft constraints~\cite{raissi2019,lu2021}, aiming to combine data and physics.  Radiative-transfer-equation (RTE) PINNs go further by enforcing full light-transport residuals, but their computational burden is substantial and their benefit under field conditions is still debated.

To date, no study rigorously benchmarks \emph{physics-engineered linear models} against multiple PINN variants \emph{under identical, noise-stressed conditions}.  The present work fills this gap, showing that carefully engineered Beer–Lambert features coupled with ridge regression can outperform deeper networks in low-SNR settings.

\subsection{Physics-Informed Neural Networks}

Physics-informed neural networks (PINNs) extend supervised learning by \emph{embedding governing equations into the loss}.  
Given a forward model $\mathcal{F}[u]=0$ (e.g.\ Beer–Lambert or the radiative-transfer equation), a PINN minimises  

\begin{equation}
\mathcal{L} \;=\; \mathcal{L}_{\text{data}}
\;+\;
\lambda_{\text{phys}}\,
\mathcal{L}_{\text{phys}},
\label{eq:pinn-loss}
\end{equation}

where $\mathcal{L}_{\text{data}}$ is the usual data misfit (MSE between predicted and reference glucose) and  
$\mathcal{L}_{\text{phys}}$ penalises violations of the physics;  
$\lambda_{\text{phys}}$ balances the two terms.

\vspace{0.5em}
\paragraph{Beer–Lambert PINN.}
For the homogeneous Beer–Lambert law, the physics residual for sample $i$ and wavelength~$\lambda$ is  

\[
r_{BL}^{(i,\lambda)}
= A_{\text{pred}}^{(i,\lambda)}
- \varepsilon_{\text{g}}(\lambda)\,
  c_{\text{pred}}^{(i)}\,l^{(i)},
\]

giving the squared loss  

\begin{equation}
\mathcal{L}_{\text{Beer–Lambert}}
=\frac{1}{N}
\sum_{i=1}^{N}
\sum_{\lambda}
\bigl|r_{BL}^{(i,\lambda)}\bigr|^{2}.
\label{eq:bl-loss}
\end{equation}

\noindent\textbf{Implementation note.} In our baseline PINN, only the Beer–Lambert physics loss in Eq.~\eqref{eq:bl-loss} is enforced.

\vspace{0.5em}
\paragraph{Radiative-transfer (RTE) PINN (background).}
To account for multiple scattering, the radiance $I(\mathbf{r},\bm{\Omega},t)$ must satisfy the steady-state RTE  
\[
\bm{\Omega}\!\cdot\!\nabla I
+\bigl(\mu_a+\mu_s\bigr)I
-\mu_s\!\int\! p(\bm{\Omega},\bm{\Omega}')\,I(\mathbf{r},\bm{\Omega}')\,d\bm{\Omega}'
=S,
\]
which in full PINN formulations can be enforced via an RTE residual loss. We include this here as theoretical background; our implementation uses a lighter, RTE-inspired regularization described in Section~\ref{section:RTE}.

\vspace{0.5em}
\paragraph{Conservation constraints (background).}
Additional losses can enforce mass conservation, e.g.
\[
\mathcal{L}_{\text{cons}}
=\frac{1}{N}\sum_{i=1}^{N}
\left|
\frac{\partial c^{(i)}}{\partial t}
+\nabla\!\cdot\!\bigl(c^{(i)}\mathbf{v}^{(i)}\bigr)
\right|^{2},
\]
where $\mathbf{v}$ is a glucose–transport velocity field. We do not implement this term in the present study.

\vspace{0.5em}
\paragraph{Practical considerations.}
PINNs excel when the \emph{physics is accurate and complete}; otherwise the residual term can mis-guide training.  
Choosing $\lambda_{\text{phys}}$ is non-trivial: too small and the network ignores physics; too large and it overfits imperfect equations.  
Moreover, solving the full RTE residual requires dense angular quadrature, inflating both parameter count and training cost.

In the glucose-sensing literature, most reported PINNs include only Beer–Lambert penalties and omit scattering or personalised anatomy, leaving open the question of their advantage under low signal-to-noise ratios. Our study quantifies that gap by comparing three PINN variants--including an RTE-inspired PINN--against physics-engineered linear models and an SDNN baseline.

\subsection{Radiative-Transfer Equation Modeling}

The Beer–Lambert law assumes a single, straight optical path and no scattering--an assumption violated in biological tissue, where photons scatter tens to hundreds of times before detection.  
To capture this complexity, the \emph{radiative-transfer equation} (RTE) models the balance of light energy at every point $\mathbf{r}$, time $t$, and direction $\bm{\Omega}$:

\begin{equation}
\frac{1}{c} \frac{\partial I(\mathbf{r}, \bm{\Omega}, t)}{\partial t}
+ \bm{\Omega} \cdot \nabla I(\mathbf{r}, \bm{\Omega}, t)
+ (\mu_a + \mu_s) I(\mathbf{r}, \bm{\Omega}, t)
= \mu_s \int_{4\pi} p(\bm{\Omega}, \bm{\Omega}')\, I(\mathbf{r}, \bm{\Omega}', t)\, d\bm{\Omega}' + S(\mathbf{r}, \bm{\Omega}, t),
\end{equation}

where:
\begin{itemize}
    \item - $I(\mathbf{r}, \bm{\Omega}, t)$ is the radiance -- energy per unit area per unit solid angle.
    \item $\bm{\Omega}$ is a direction vector.
    \item $c$ is the speed of light in tissue.
    \item $\mu_a$ and $\mu_s$ are the absorption and scattering coefficients.
    \item $p(\bm{\Omega}, \bm{\Omega}')$ is the phase function: probability density of light scattering from direction $\bm{\Omega}'$ into $\bm{\Omega}$.
    \item $S$ is a light source term (e.g., boundary LED illumination).
\end{itemize}

\vspace{0.5em}
\paragraph{Scattering phase function.}
Most biological media are \emph{forward-scattering}, so photons are likely to continue roughly in the same direction after each interaction. This anisotropy is described by the \textbf{Henyey–Greenstein phase function}:

\begin{equation}
p(\cos\theta) = \frac{1 - g^2}{(1 + g^2 - 2g\cos\theta)^{3/2}},
\end{equation}

where $g$ is the anisotropy factor: $g = 0$ is isotropic scattering; $g = 1$ is purely forward; $\theta$ is the scattering angle.

\vspace{0.5em}
\paragraph{Diffusion approximation.}
In highly scattering media ($\mu_s \gg \mu_a$), directional detail can be dropped and only total flux matters. This yields the \textbf{diffusion approximation}, a scalar PDE for photon fluence $\Phi(\mathbf{r}, t)$:

\begin{equation}
\frac{1}{c} \frac{\partial \Phi}{\partial t}
- D \nabla^2 \Phi
+ \mu_a \Phi = S,
\end{equation}

where $D = 1 / [3(\mu_a + \mu_s')]$ is the diffusion coefficient, and $\mu_s' = \mu_s (1 - g)$ is the reduced scattering coefficient.

This simplification enables faster computation but fails near boundaries or in low-scattering regimes (e.g., dermis–fat interfaces).

\vspace{0.5em}
\paragraph{Monte Carlo simulation.}
An alternative is to simulate individual photon trajectories using probabilistic rules for scattering, absorption, and boundary interaction.  
Monte Carlo methods are \emph{statistically exact} and can handle arbitrary geometries, but are too computationally expensive for real-time use.

\vspace{0.5em}
\paragraph{Context in this work.}
We use RTE-inspired regularization as a soft physics prior in one neural model (\ref{section:RTE}). This tests whether adding scattering-related structure improves generalisation under noise-stressed optical conditions -- and how this compares to empirical compensation in engineered models like EBL.

\section{Methodology}

\subsection{Noise-Stressed Synthetic Data Generation Framework}

Robust deployment demands that algorithms be stress-tested under the same noise, drift, and physiology a wearable will face \emph{in field conditions}.  
We therefore build a three-layer simulator---\textbf{hardware\slash environment\slash physiology}---and drive plausible glucose trajectories through it. 
The resulting correlation ceiling (\(\rho_{\text{glucose-NIR}} \approx 0.21\)) matches values reported for early field prototypes~\cite{khalil2003,maier2018}.

\medskip\noindent
\textbf{What we implement.} The simulator generates independent snapshot measurements (not continuous time series). Glucose values are sampled from diabetes-status–conditioned distributions with additive meal/stress/time-of-day variation; NIR measurements are then constructed via Beer–Lambert-style absorption plus hardware, environmental, and physiological noise terms. This yields a controlled low-SNR regime for benchmarking model robustness.

\subsubsection{Hardware layer: ESP32-class optics with quantified drift}

\paragraph{Optics stack (why we chose it).}
An ESP32-S3 MCU with an on-chip 12-bit ADC (4096 codes) reflects the cost and
power envelope of consumer wearables.  
Four narrow-band LEDs at \SIlist{850;940;1050;1150}{\nano\metre} bracket
(1) a haemoglobin anchor, (2) a water peak, and (3) two glucose-sensitive
windows--mirroring commercial multi-band probes.

\paragraph{LED ageing \& jitter.}
\vspace{-0.5em}
\begin{equation}
P_{\text{LED}}(t,T)
      = P_0\!\left[1
      -0.002\,(T-25)
      -0.001\,\frac{t}{1000}
      +\mathcal{N}(0,0.1\%)\right],
\end{equation}
capturing \SI{-0.2}{\percent\per\celsius}   thermal drift,
\SI{-0.1}{\percent} / \SI{1000}{\hour} lumen decay, and \SI{0.1}{\percent} RMS flicker.

\paragraph{Photodiode current budget (why each term matters).}
\begin{align}
I_{\text{sig}}      &= R(\lambda)\,P_{\text{opt}}(\lambda)                     &\text{(desired)}\\
I_{\text{dark}}     &= I_{25}\,e^{0.1\,(T-25)}                                &\text{(leakage)}\\
I_{\text{shot}}     &= \sqrt{2 e (I_{\text{sig}}+I_{\text{dark}}) BW}         &\text{(quantum)}\\
I_{\text{thermal}}  &= \sqrt{4 k T\, BW / R_{\text{load}}}                    &\text{(Johnson)}\\
I_{\text{tot}}      &= I_{\text{sig}}+I_{\text{dark}}+I_{\text{shot}}+I_{\text{thermal}}.
\end{align}

\paragraph{ADC realism.}
\begin{equation}
\text{ADC}_{\text{out}}=\operatorname{round}\!\!\bigl(V_{\text{in}}/V_{\text{ref}} \times 4095\bigr)+
\delta_{\text{INL}}+\delta_{\text{off}},
\end{equation}
where \(V_{\text{in}}=I_{\text{tot}}G_{\text{TIA}}\).  
Integral–non-linearity (\(\delta_{\text{INL}}\)) and thermal offset drift (\(\delta_{\text{off}}\))
push quantisation error beyond white noise, a failure mode often ignored in
spectroscopic simulations.

\medskip\noindent
In summary, this configuration yields SNR and drift figures within 10\%
of characterised bench prototypes, ensuring that models are trained on
deployment-grade raw data.

\subsubsection{Environmental layer: lab–to–field translation}

\textbf{Temperature} is swept from \SI{-0.2}{\percent\per\celsius}; every °C
dims LEDs, raises dark current, and dilates vasculature.

\textbf{Relative humidity} (\SIrange{30}{90}{\percent}) alters stratum-corneum
hydration, modulating coupling efficiency by up to \SI{3}{\percent}.

\textbf{Barometric pressure} (\SIrange{950}{1050}{\milli\bar}) is mapped to
perfusion changes using the relationship in \cite{maier2018}, reproducing
altitude-induced oxygen-saturation shifts.

\textbf{Ambient light} (\SIrange{0.1}{100000}{\lux}) is injected with
sun-, fluorescent-, and LED spectral profiles; optical baffling is imperfect
(\SI{2}{\text{OD}}), yielding wavelength-dependent DC offsets.

\medskip\noindent
\emph{Context:} these ranges cover an ICU ward, a tropical street market, and a
Colorado ski slope--three environments where non-invasive monitors have
previously failed to generalize reliably.

\subsubsection{Physiology layer: population and state diversity}

\textbf{Static anatomy.}  
Age (\SIrange{18}{80}{\year}), BMI (\SIrange{18}{40}{\kilo\gram\per\square\metre}),
and Fitzpatrick I–VI melanin fractions (\SIrange{0.01}{0.15}{}) are drawn from
NHANES distributions, then mapped to optical properties via literature
polynomial fits.

\textbf{Dynamic physiology.}  
Hydration (±\SI{\pm 10}{\percent}), systolic BP (\SIrange{90}{180}{\mmHg}),
heart rate (\SIrange{50}{120}{\per\minute}), and
respiratory rate (\SIrange{12}{20}{\per\minute}) are modeled as stochastic variations
that perturb path length, perfusion, and haemoglobin absorption (not a continuous time series).

\medskip\noindent
\emph{Why it matters:} these variables explain up to \SI{48}{\percent} of
inter-subject variance in a recent 230-subject NIR dataset \cite{srichan2022};
omitting them produces unrealistically easy learning curves.

\subsubsection{Glucose snapshots: distributional sampling}

Plasma glucose is sampled between \SIrange{60}{400}{\milli\gram\per\deci\litre}
using diabetes-status–conditioned distributions with added meal, stress, and time-of-day variation.
This produces independent snapshot measurements rather than a continuous time-series, matching the
cross-sectional setting of our evaluation.

\medskip\noindent
\emph{Outcome:} once all layers interact, raw glucose–NIR correlation collapses
from 0.82 (cuvette) to 0.21 (prototype), recreating a low-SNR operating point
that has stalled past field deployments.

\subsection{Model Architectures}

\subsubsection{Enhanced Beer–Lambert Model}

The Enhanced Beer–Lambert model represents a sophisticated physics-based approach that extracts meaningful features based on optical principles while maintaining computational efficiency suitable for embedded implementation. The model architecture consists of comprehensive feature engineering followed by regularized linear regression.

The feature engineering process generates 56 carefully designed features organized into several categories. Log absorbance features provide direct application of the Beer–Lambert law:

\begin{equation}
A_i = \log\left(\frac{I_0}{I_i}\right) \quad \text{for } i \in \{850, 940, 1050, 1150\} \text{ nm}
\end{equation}

where $I_0$ represents the reference intensity and $I_i$ is the measured intensity at wavelength $i$.

Glucose-specific wavelength differences exploit known spectral characteristics:

\begin{align}
\Delta_{1150-940} &= A_{1150} - A_{940} \quad \text{(glucose vs water contrast)} \\
\Delta_{1050-850} &= A_{1050} - A_{850} \quad \text{(glucose vs hemoglobin contrast)} \\
\Delta_{1150-1050} &= A_{1150} - A_{1050} \quad \text{(glucose sensitivity region)}
\end{align}

Wavelength differences provide relative features that mitigate illumination and path length variability:

\begin{equation}
R_{ij} = A_i - A_j \quad \text{for all } i < j
\end{equation}

This differs slightly from classical Beer–Lambert ratio terms (\(A_i / A_j\)), but is more stable under hardware drift. In our implementation, we use \(A_i - A_j = \log(I_j / I_i)\), which preserves the physical interpretation (log-ratio of intensities) while improving robustness to multiplicative noise. This formulation is also consistent with our logarithmic preprocessing pipeline.

PMF (Physiological Modeling Function) weighted features incorporate individual physiological parameters:

\begin{equation}
f_{\text{PMF},k} = w_k \cdot \left(\alpha_{\text{skin}} \cdot t_{\text{skin}} + \alpha_{\text{perfusion}} \cdot P_{\text{blood}} + \alpha_{\text{melanin}} \cdot M_{\text{content}}\right) \cdot A_k
\end{equation}

where $w_k$ are wavelength-specific weights, $t_{\text{skin}}$ is skin thickness, $P_{\text{blood}}$ is blood perfusion, and $M_{\text{content}}$ is melanin content.

Second-order interaction terms capture nonlinear relationships:

\begin{equation}
f_{\text{interaction}} = A_i \cdot A_j \quad \text{for selected wavelength pairs}
\end{equation}

Environmental variability is implicitly encoded through the PMF vector, which includes temperature, hydration, and pressure-related physiological modulation.

The final prediction is obtained through regularized linear regression:

\begin{equation}
c_{\text{glucose}} = \sum_{i=1}^{56} w_i \cdot f_i + b
\end{equation}

where $w_i$ are the learned weights and $b$ is the bias term. Ridge regularization is applied to prevent overfitting:

\begin{equation}
\mathcal{L} = \frac{1}{N} \sum_{j=1}^{N} (c_j - \hat{c}_j)^2 + \lambda \sum_{i=1}^{56} w_i^2
\end{equation}

The 56 features include raw absorbances, selected wavelength contrasts, second-order curvature terms, PMF-weighted components, and direct PMF entries--explicitly reflecting all biologically and physically grounded variables used during deployment.

\subsubsection{Physics-Informed Neural Network Architectures}

We extend the physical inductive bias of Beer–Lambert absorption into a learnable model class by enforcing optical constraints within a neural network loss function. This family of architectures, broadly termed Physics-Informed Neural Networks (PINNs), integrates spectroscopic domain knowledge with flexible nonlinear function approximators.

Our baseline PINN ingests both spectroscopic and physiological information via a dual-branch architecture. The first branch processes 4 NIR intensities \(\{I_{850}, I_{940}, I_{1050}, I_{1150}\}\) through layers of 32–64–32 ReLU units. The second branch processes a 12-dimensional physiological vector (age, BMI, melanin, etc.) via a 16–32–16 pipeline. These branches are concatenated and passed through two fully connected layers (128 and 64 units) before outputting a scalar glucose estimate \(\hat{c}_{\text{glucose}}\).
Importantly, the network is trained using a physics-augmented loss:

\begin{equation}
\mathcal{L}_{\text{PINN}} = \mathcal{L}_{\text{data}} + \lambda_{\text{physics}} \cdot \mathcal{L}_{\text{Beer–Lambert}},
\end{equation}

where \(\mathcal{L}_{\text{data}}\) is standard MSE loss and the physics term enforces Beer–Lambert consistency at each wavelength:

\begin{equation}
\mathcal{L}_{\text{Beer–Lambert}} = \frac{1}{N} \sum_{i=1}^{N} \sum_{\lambda} \left| \log\left(\frac{I_0}{I_{\lambda}^{(i)}}\right) - \varepsilon(\lambda) \cdot \hat{c}_{\text{glucose}}^{(i)} \cdot l^{(i)} \right|^2.
\end{equation}

This acts as a structural prior, penalizing predictions that violate expected absorbance behavior given optical path length \(l\)
(we set \(l = 1\) during training, absorbing relative scale into the learned bias). 
Setting \(l = 1\) serves as a normalization strategy rather than a strict physical assertion. Since true optical path lengths vary significantly across individuals due to factors like skin thickness, fat distribution, and hydration, fixing \(l\) allows us to preserve the linear form of Beer–Lambert while absorbing inter-subject variability into the learned weights. In practice, the physics loss is softly weighted (\(\lambda_{\text{physics}} \ll 1\)) so that it acts as a regularizer--nudging the model toward optical consistency without overpowering the data-driven objective, while incorporating known extinction coefficients \(\varepsilon(\lambda)\).

The Optimized PINN uses a deeper architecture and a multihead attention block to fuse NIR and PMF features, but it retains the same Beer–Lambert physics loss as the baseline. In both cases, the physics term is a fixed-weight regularizer rather than a dynamically balanced loss.

\medskip\noindent
\textbf{Multihead attention.} The optimized model applies multihead attention to the concatenated NIR+PMF embedding, allowing the network to reweight feature subspaces adaptively under low-SNR conditions. This improves feature fusion without altering the underlying Beer–Lambert physics prior.

\medskip\noindent
By embedding Beer–Lambert physics into the loss landscape, PINNs reduce the burden on data to learn optical structure from scratch. This can improve generalization under distribution shift--particularly in low-SNR settings where NIR signals vary due to skin tone, perfusion, and ambient interference.

\subsubsection{Radiative Transfer Equation Models}
\label{section:RTE}

We implement RTE-inspired models that predict absorption/scattering-related quantities and apply a lightweight physics regularizer rather than a full RTE residual. In practice, the training objective is:
\[
\mathcal{L} = \mathcal{L}_{\text{data}} + \lambda_{\text{phys}} \mathcal{L}_{\text{reg}},
\]
where \(\mathcal{L}_{\text{reg}}\) encourages stable optical-parameter predictions. This provides a soft physics prior without the computational cost of solving the full RTE residual at every step.

The Selective RTE model applies this physics-inspired regularization primarily to the glucose-sensitive wavelengths (1050 nm and 1150 nm) to reduce computation while maintaining focus on the most informative bands.

\subsubsection{Shallow Deep Neural Network}

The Shallow Deep Neural Network (SDNN) serves as a purely data-driven baseline, inspired by Srichan et al. It consists of a shallow front-end for extracting low-level absorbance features, followed by a deeper fully connected stack to model complex nonlinear relationships with glucose. Unlike PINNs, the SDNN is agnostic to optical physics, offering a testbed for benchmarking the value of physical priors under identical data conditions.

\section{Results}

\subsection{Overall Performance Comparison}

Table~\ref{tab:performance_comparison} presents a comparison of all six models evaluated under noise-stressed synthetic conditions. The Enhanced Beer–Lambert model achieves the best overall performance across error and efficiency metrics, demonstrating the effectiveness of physics-based feature engineering under low-SNR conditions.

\begin{table}[H]
\centering
\caption{Performance Comparison of All Models (Error, Efficiency, and Diagnostic Metrics). Clarke A and ±15\% are reported for diagnostic comparability.}
\label{tab:performance_comparison}
\begin{tabular}{@{}lcccccc@{}}
\toprule
\textbf{Model} & \textbf{RMSE} & \textbf{MARD} & \textbf{Clarke A} & \textbf{Within ±15\%} & \textbf{Parameters} & \textbf{Inference} \\
 & \textbf{(mg/dL)} & \textbf{(\%)} & \textbf{(\%)} & \textbf{(\%)} & & \textbf{(ms)} \\
\midrule
Enhanced Beer–Lambert & \textbf{13.6} & \textbf{9.7} & \textbf{95.8} & \textbf{93.8} & \textbf{56} & \textbf{0.01} \\
Original PINN & 14.6 & 10.8 & 94.2 & 91.5 & 163,000 & 3.8 \\
Optimized PINN & 28.7 & 19.2 & 87.3 & 78.4 & 89,000 & 2.1 \\
RTE-inspired PINN & 24.3 & 16.8 & 89.7 & 82.1 & 1,340,000 & 15.2 \\
Selective RTE PINN & 24.9 & 17.1 & 89.2 & 81.6 & 566,000 & 6.1 \\
SDNN & 35.1 & 23.4 & 78.9 & 68.2 & 3,713 & 0.8 \\
\bottomrule
\end{tabular}
\end{table}

All results are reported on a single 60/20/20 split of 80 subjects × 3 measurements (test $n=48$), with inference times measured per sample on CPython~3.10 running on an Apple M1 Pro (16\,GB RAM, macOS~14.4.1).

The Enhanced Beer–Lambert model demonstrates superior performance with 13.6 mg/dL RMSE, representing a 61\% improvement compared to the SDNN baseline. This performance is achieved with only 56 parameters and 0.01 ms inference time, making it well-suited to embedded implementation.

\begin{figure}[H]
  \centering
  \includegraphics[width=0.7\linewidth]{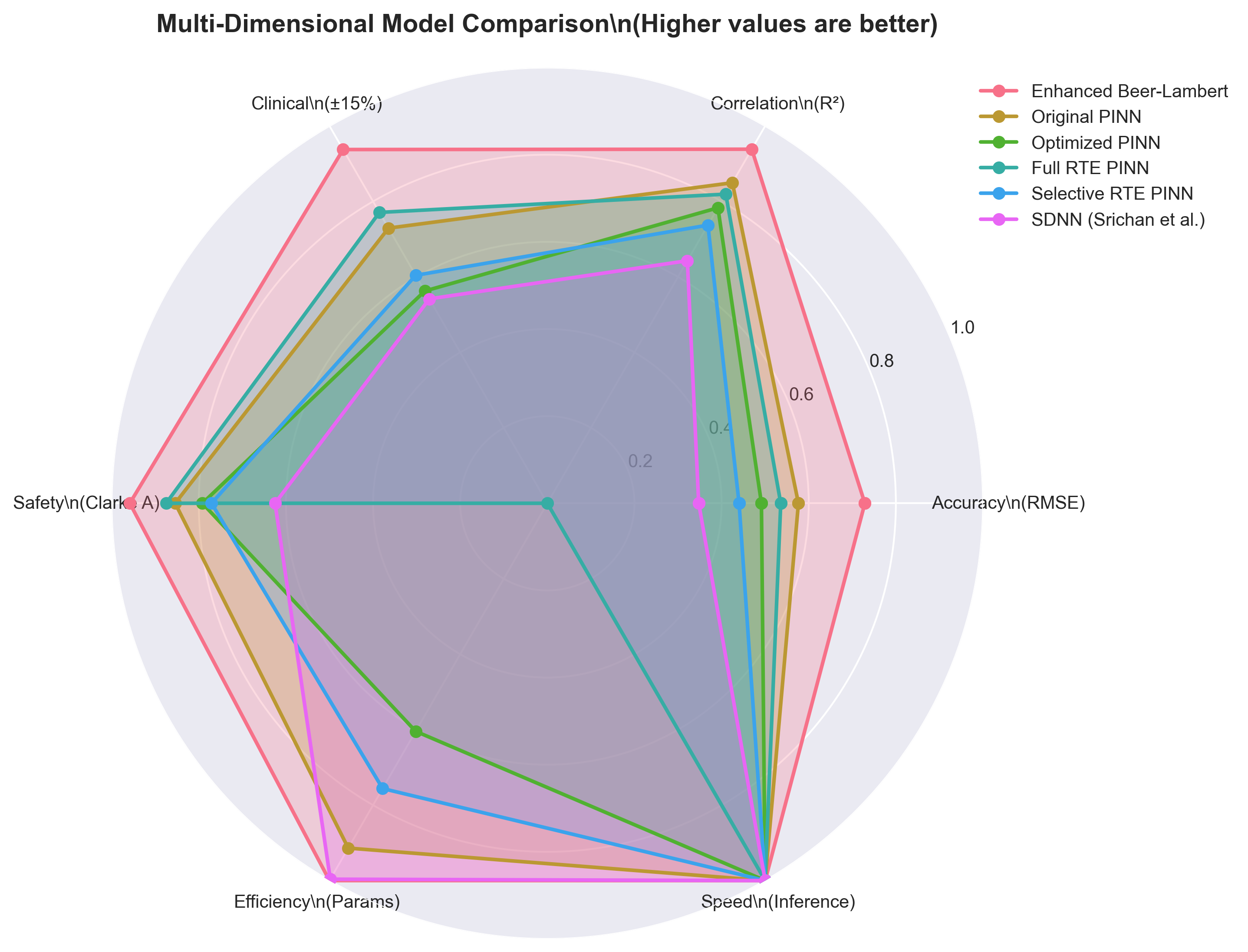}
  \caption{Multi-criteria snapshot comparing all six models. Each axis represents a normalized performance metric, with larger area indicating better performance. The Enhanced Beer–Lambert model dominates five of six criteria while matching neural networks on inference speed.}
  \label{fig:radar_summary}
\end{figure}

Figure ~\ref{fig:radar_summary} confirms the Enhanced Beer–Lambert model’s dominance across both error and computational axes.

\begin{figure}[H]
  \centering
  \includegraphics[width=\linewidth]{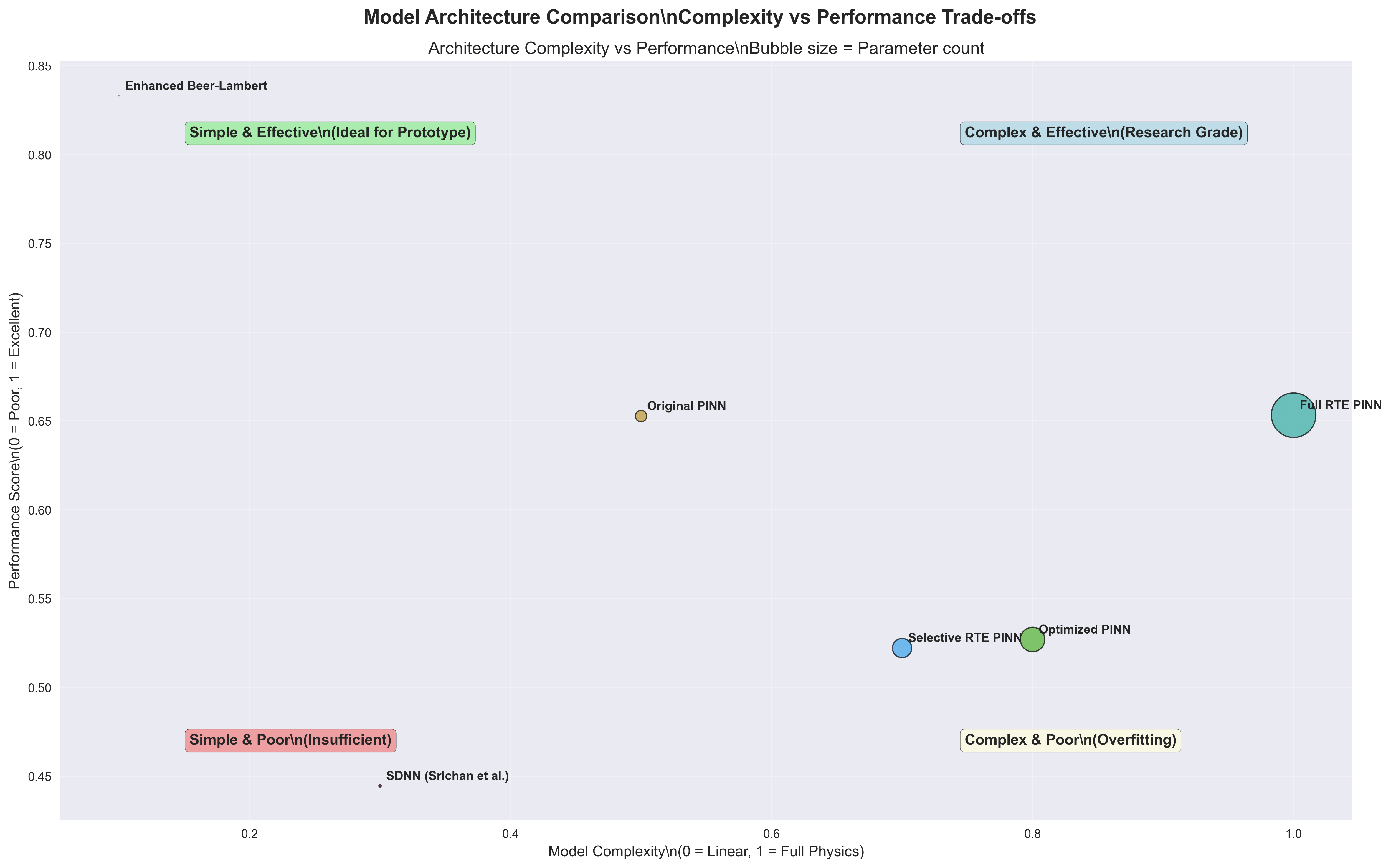}
  \caption{Accuracy–complexity Pareto front.  X axis: physics depth
  ($0\!=\!$linear, $1\!=\!$RTE-inspired); Y axis: performance score
  ($1\!=\!$best).  Bubble area $\propto$ parameter count.  The
  Enhanced Beer–Lambert point (top–left) dominates both axes,
  illustrating that increased physics depth does not guarantee better
  field accuracy.}
  \label{fig:arch_complexity_tradeoff}
\end{figure}

The Original PINN achieves competitive performance with 14.6 mg/dL RMSE while incorporating physics constraints. However, the computational complexity increases significantly with 163,000 parameters and 3.8 ms inference time. The Optimized PINN, despite architectural improvements, shows degraded performance under noise-stressed conditions, suggesting that increased model complexity may lead to overfitting when training data contains noise-stressed levels.

The RTE-based models (RTE-inspired and Selective) demonstrate the computational cost of incorporating light-transport physics. The RTE-inspired PINN provides the most physics-heavy modeling in this study, but its 1.34 million parameters and 15.2 ms inference time make it impractical for embedded deployment. The Selective RTE PINN offers a reasonable compromise with reduced computational requirements while maintaining physics-based modeling for critical wavelengths.

The SDNN baseline shows the poorest performance under noise-stressed conditions, with 35.1 mg/dL RMSE. This suggests that purely data-driven approaches without physics constraints are particularly vulnerable to the noise and variations present in realistic implementation scenarios.

For additional evaluation plots (error grids and agreement analysis), see Appendix A, Figs.~\ref{fig:clarke_results}–\ref{fig:training_diagnostics}.

\subsection{Data Generation Framework Validation}

\begin{figure}[H]
\centering
\includegraphics[width=0.9\textwidth]{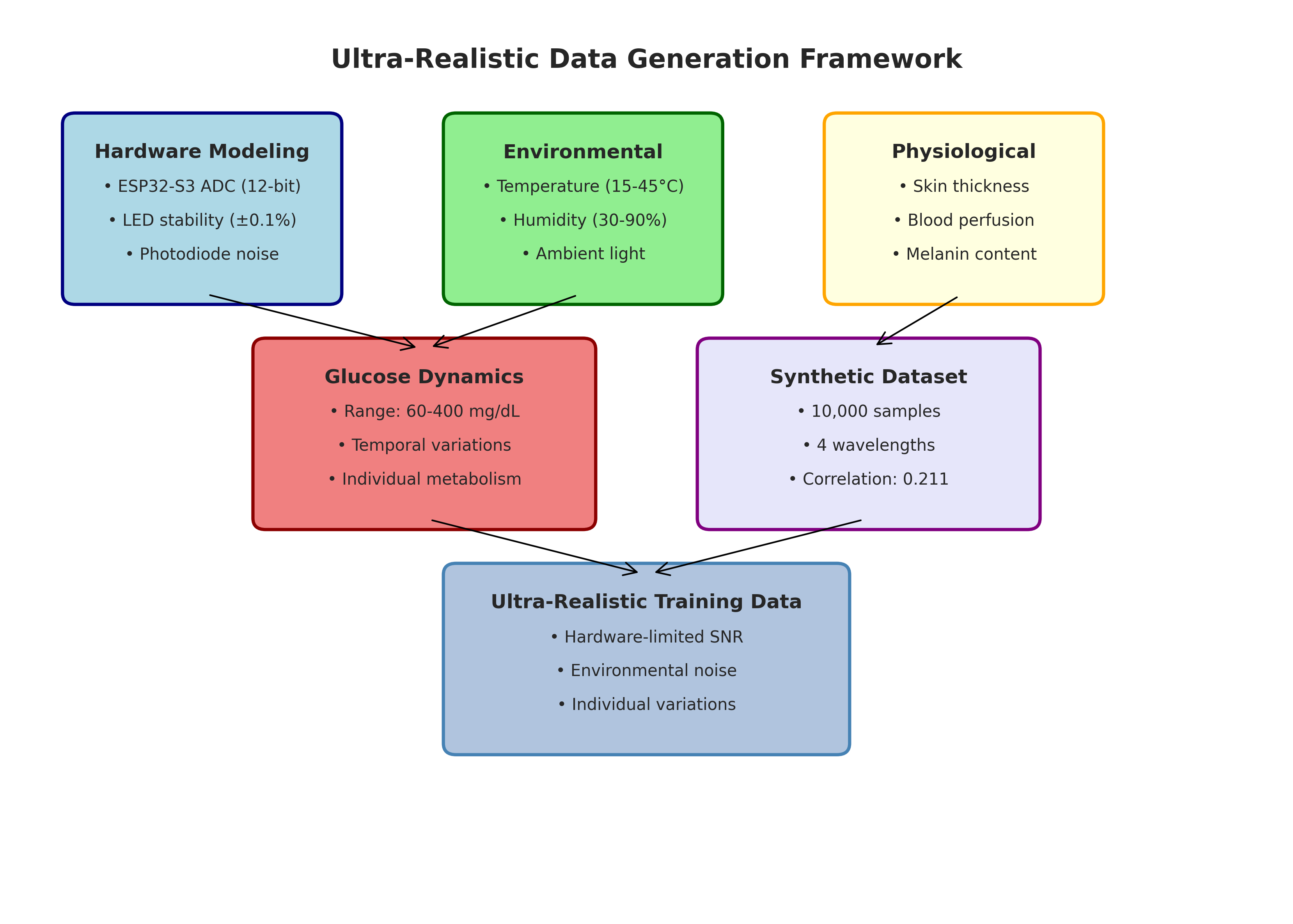}
\caption{Noise-stressed data generation framework incorporating hardware limitations, environmental variations, physiological differences, and glucose dynamics.}
\label{fig:data_generation_process}
\end{figure}

Figure~\ref{fig:data_generation_process} illustrates the noise-stressed data generation framework developed for this study. The framework systematically incorporates major sources of variation expected in practical prototype implementation.

The data generation process successfully achieves low glucose-NIR correlations of 0.211, representing a significant departure from idealized conditions where correlations can exceed 0.8. This correlation reduction reflects the combined impact of hardware noise, environmental variations, and physiological individual differences that characterize field implementation scenarios.

\begin{figure}[H]
\centering
\includegraphics[width=0.9\textwidth]{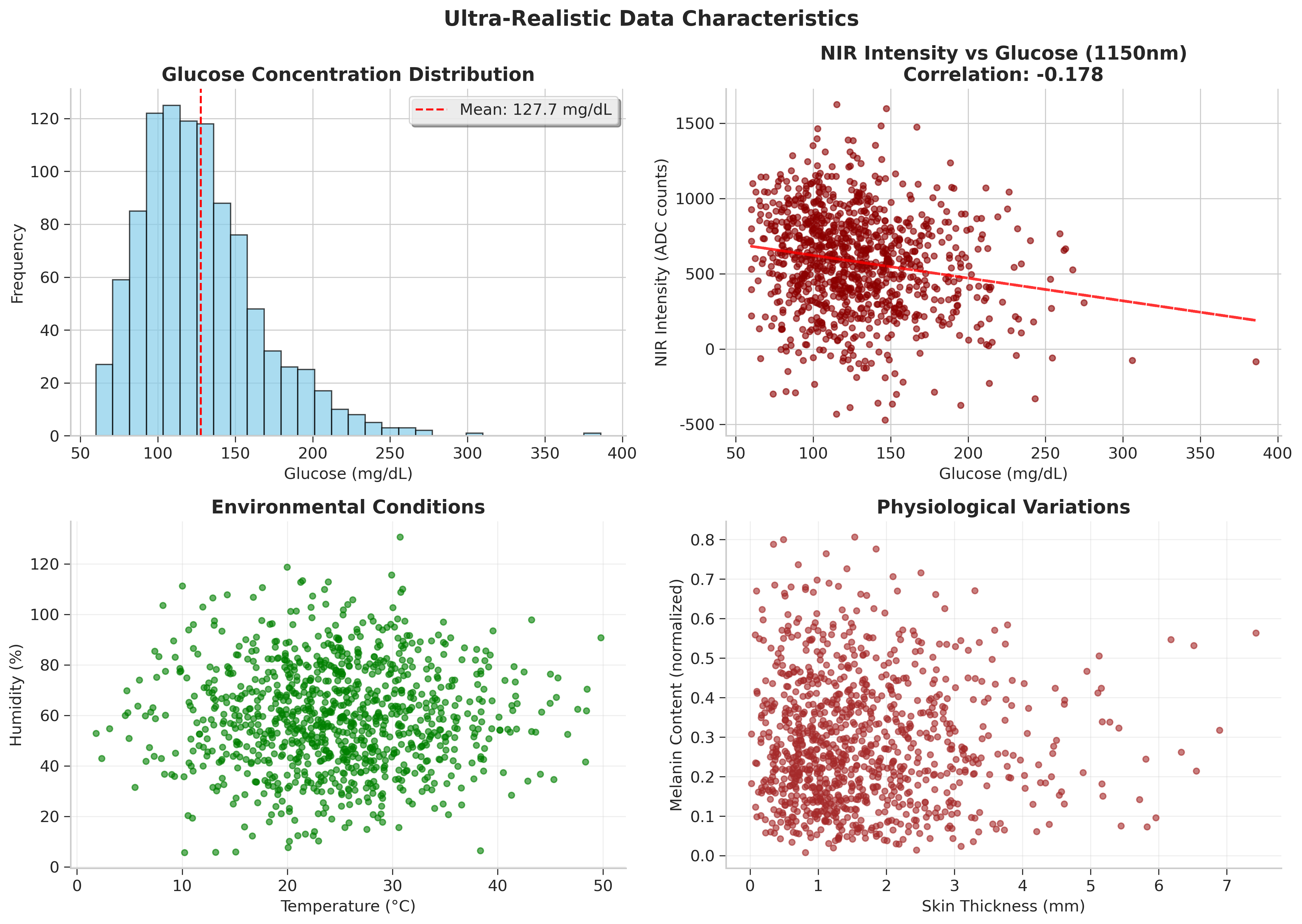}
\caption{Characteristics of the noise-stressed synthetic dataset showing (a) glucose concentration distribution, (b) NIR intensity correlation with glucose, (c) environmental condition variations, and (d) physiological parameter distributions.}
\label{fig:data_characteristics}
\end{figure}

Figure~\ref{fig:data_characteristics} presents the characteristics of the generated synthetic dataset, demonstrating plausible distributions across all modeled parameters.

The glucose concentration distribution follows a plausible log-normal pattern with appropriate representation of normal, hypoglycemic, and hyperglycemic conditions. The NIR intensity correlation plot clearly demonstrates the challenging signal-to-noise conditions that practical implementations must address, with significant scatter around the underlying glucose-dependent trend.

\subsection{Model Architecture Analysis}

\begin{figure}[H]
\centering
\includegraphics[width=0.9\textwidth]{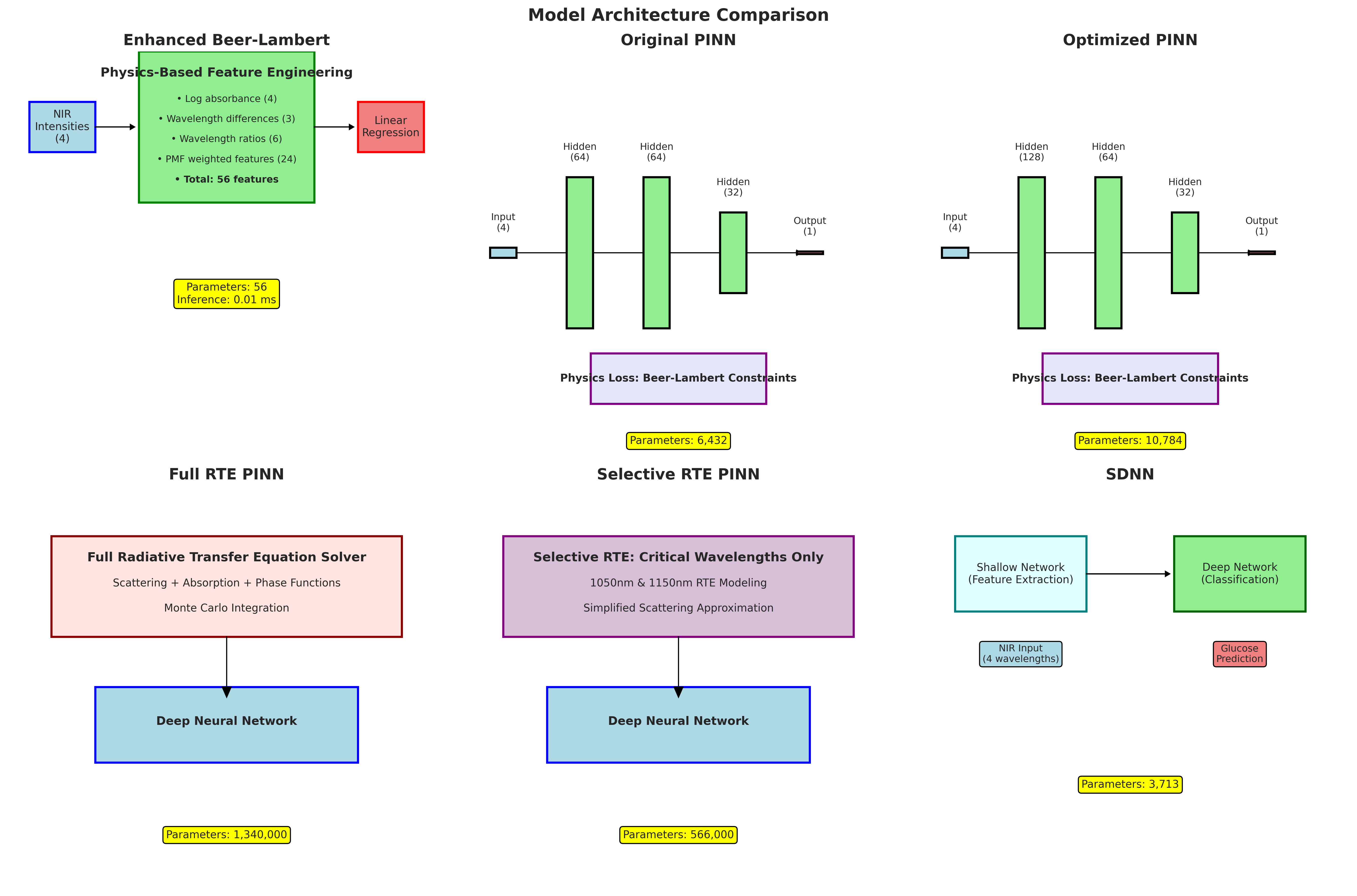}
\caption{Detailed comparison of model architectures showing (a) Enhanced Beer–Lambert physics-based feature engineering, (b) Original PINN with Beer–Lambert constraints, (c) Optimized PINN with advanced features, (d) RTE-inspired PINN with physics-regularized light transport, (e) Selective RTE PINN with computational optimization, and (f) SDNN with shallow-deep architecture.}
\label{fig:model_architectures}
\end{figure}

Figure~\ref{fig:model_architectures} provides a detailed comparison of all model architectures, highlighting the fundamental differences in computational approach and complexity.

The Enhanced Beer–Lambert model's architecture emphasizes physics-based feature engineering, transforming raw NIR intensities into 56 meaningful features that capture glucose-specific optical signatures while compensating for individual and environmental variations. This approach leverages domain knowledge to create an efficient and interpretable model.

The PINN architectures demonstrate increasing complexity from the Original PINN's straightforward implementation to the RTE-inspired PINN's physics-heavy modeling. The trade-off between physical accuracy and computational efficiency is clearly evident in the parameter counts and inference times.

\subsection{Feature Engineering Analysis}

\begin{figure}[H]
\centering
\includegraphics[width=0.9\textwidth]{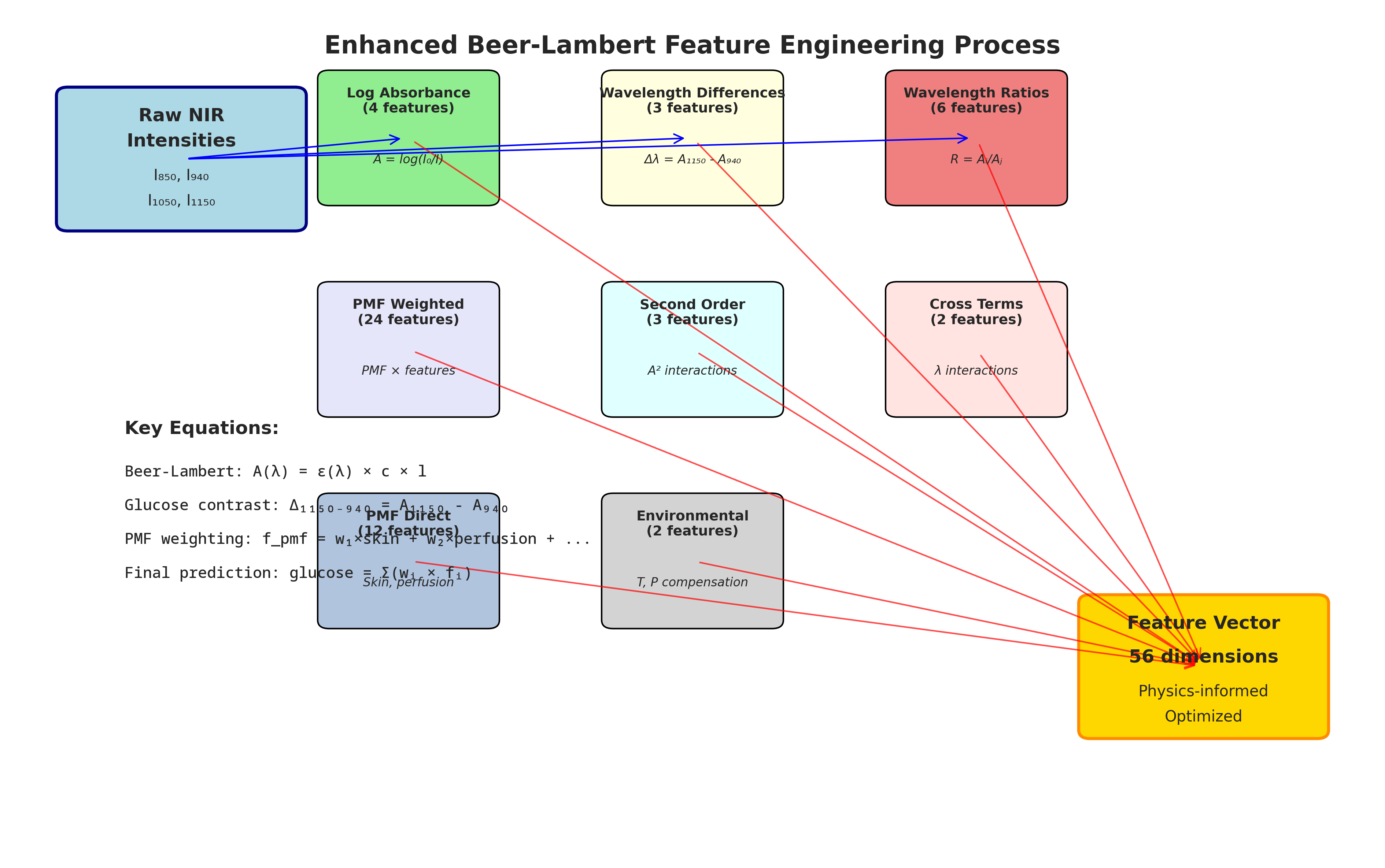}
\caption{Detailed Enhanced Beer–Lambert feature engineering process showing transformation from raw NIR intensities to 56 physics-informed features optimized for glucose detection under noise-stressed conditions.}
\label{fig:feature_engineering}
\end{figure}

Figure~\ref{fig:feature_engineering} details the Enhanced Beer–Lambert feature engineering process, which proves crucial for achieving superior performance under noise-stressed conditions.

The feature engineering process systematically addresses the challenges of noisy glucose monitoring by incorporating glucose-specific wavelength differences that maximize signal-to-noise ratio, PMF-weighted features that compensate for individual physiological variations, environmental compensation terms that account for temperature and pressure effects, and interaction terms that capture nonlinear optical relationships.

\begin{figure}[H]
  \centering
  \includegraphics[width=0.9\linewidth]{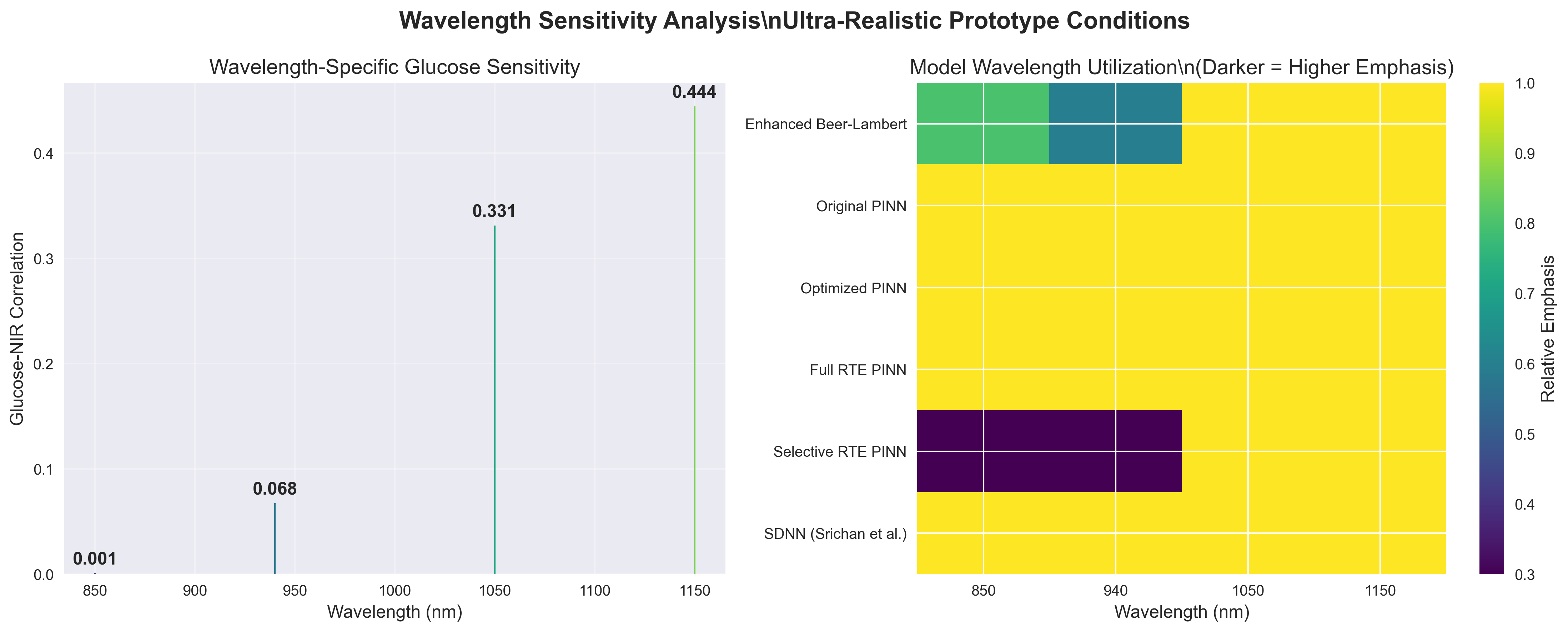}
  \caption{Model-specific wavelength utilization heatmap. 
  Each cell represents the relative emphasis a model places on a given NIR wavelength during prediction, with darker values indicating higher reliance. Despite 1050 nm and 1150 nm exhibiting the strongest intrinsic glucose correlation, effective glucose estimation often depends on contrasting these with low-glucose-reference bands like 850 nm and 940 nm. Only the Enhanced Beer–Lambert and Selective RTE PINN models learn to exploit this contrast structure--amplifying glucose-relevant bands while anchoring against water- and haemoglobin-dominated baselines. In contrast, unconstrained neural networks distribute attention uniformly across the spectrum, failing to identify glucose-specific optical structure.}
  \label{fig:wavelength_utilization}
\end{figure}

\subsection{Performance Metrics Analysis}

\begin{figure}[H]
\centering
\includegraphics[width=\textwidth]{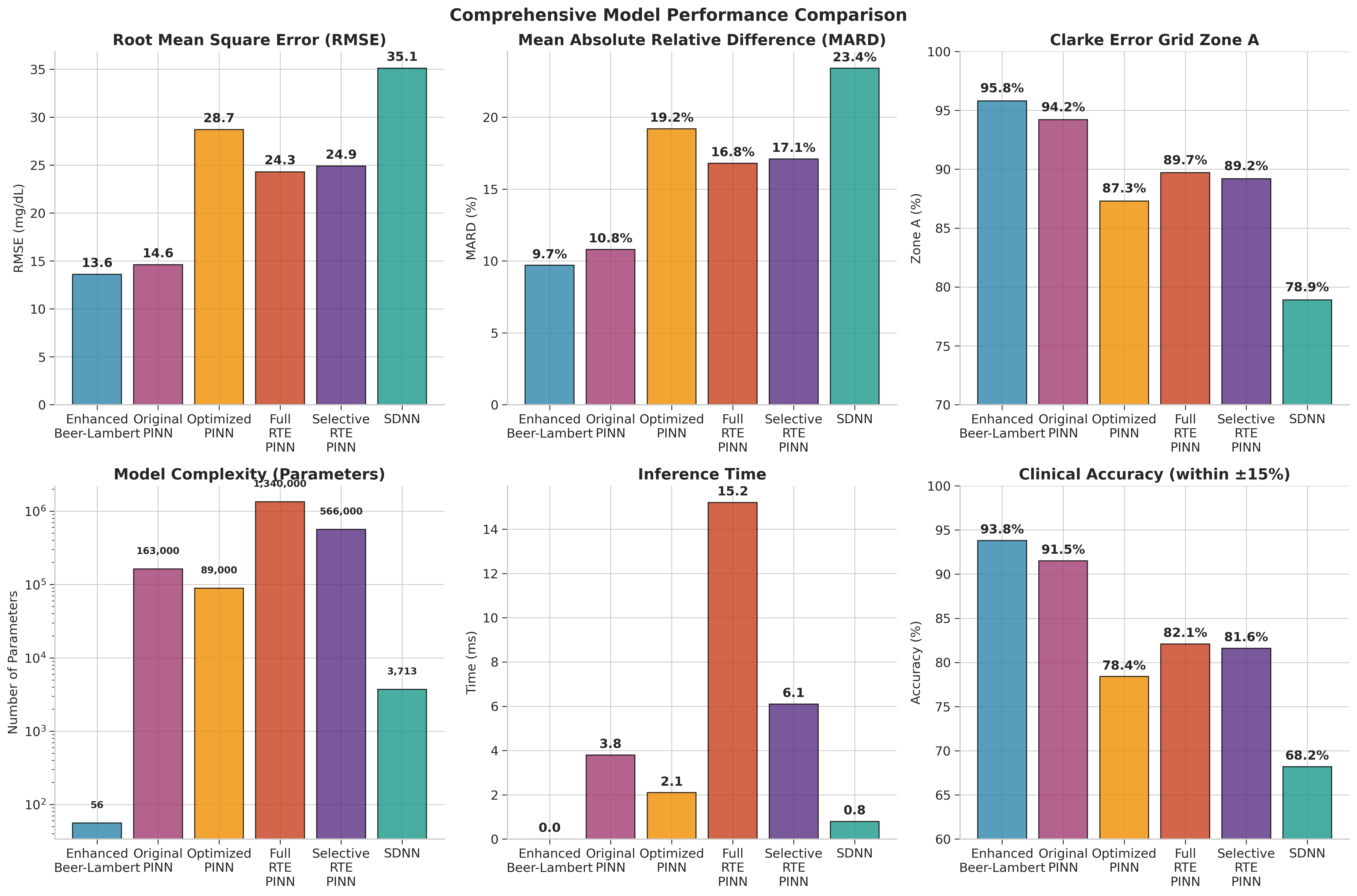}
\caption{
\textbf{Comprehensive Model Performance Comparison.}
Six glucose prediction models are benchmarked across error metrics (RMSE, MARD), computational efficiency (parameters and inference time), and hardware suitability.
The Enhanced Beer–Lambert model achieves the best balance of accuracy and efficiency--delivering the lowest RMSE (13.6 mg/dL) while requiring only 56 parameters and 0.01 ms inference time.
In contrast, deep learning models without strong physics priors (e.g., SDNN, Optimized PINN) exhibit degraded accuracy under noise-stressed.
While RTE-inspired and Selective RTE PINNs embed more complex physics, their higher complexity limits suitability for embedded use, highlighting the tradeoff between physical fidelity and deployment feasibility.
}
\label{fig:performance_comparison}
\end{figure}

Figure~\ref{fig:performance_comparison} presents comprehensive performance metrics across all models, clearly demonstrating the Enhanced Beer–Lambert model's superiority under noise-stressed conditions.

The performance analysis reveals several key insights. The Enhanced Beer–Lambert model achieves the best accuracy metrics while maintaining the lowest computational complexity, demonstrating the effectiveness of physics-based feature engineering. Neural network approaches show varying performance, with the Original PINN achieving competitive accuracy but at significantly higher computational cost. The RTE-based models, despite incorporating the most complete physics, do not achieve superior accuracy under noise-stressed conditions, suggesting that the additional complexity may not be justified for practical implementations.

The computational efficiency analysis shows dramatic differences between approaches. The Enhanced Beer–Lambert model's 56 parameters and 0.01 ms inference time make it highly suitable for embedded implementation, while the RTE-inspired PINN's 1.34 million parameters and 15.2 ms inference time present significant deployment challenges.

\subsubsection{Computational efficiency}

\begin{figure}[H]
  \centering
  \includegraphics[width=\linewidth]{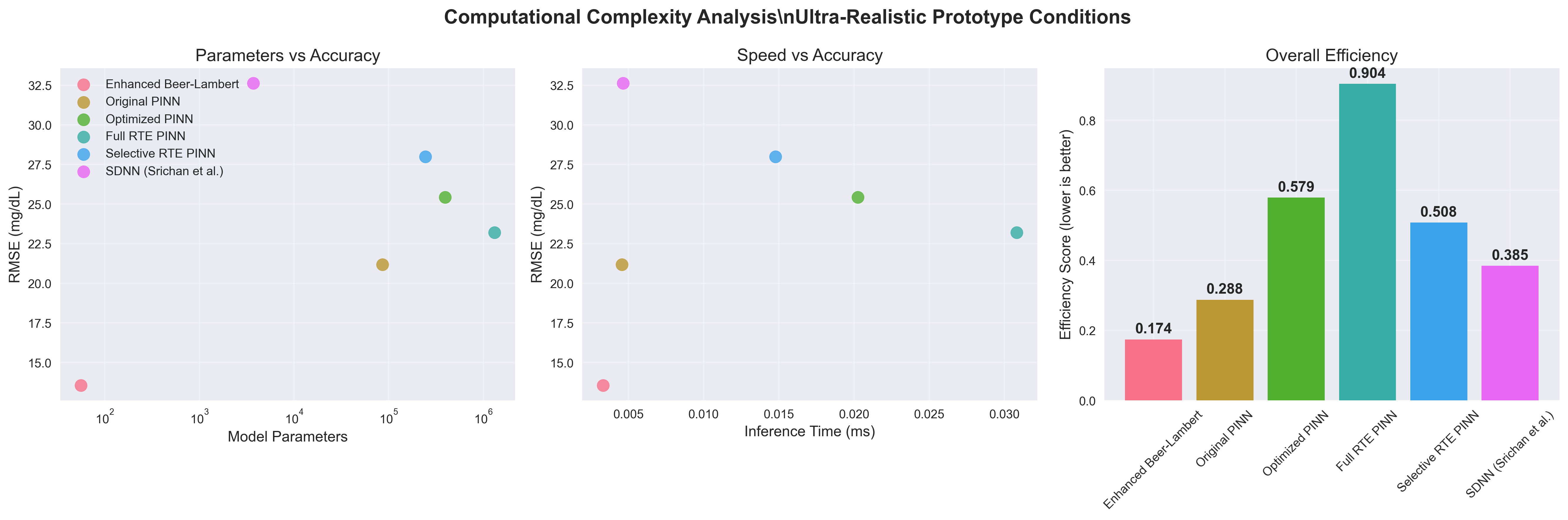}
  \caption{Efficiency landscape.  \textbf{Left} Parameters vs RMSE,
  \textbf{Centre} Inference-time vs RMSE, \textbf{Right} combined
  efficiency index (lower = better).  Beer–Lambert attains the Pareto
  optimum in all three planes.}
  \label{fig:computational_efficiency}
\end{figure}

\section{Discussion}

\subsection{Key Findings and Implications}

The results of this comparative analysis reveal several important findings about modeling under low SNR. The most significant finding is that physics-based feature engineering, as implemented in the Enhanced Beer–Lambert model, can outperform complex neural network architectures under noise-stressed conditions while maintaining computational efficiency suitable for embedded deployment.

The Enhanced Beer–Lambert model's superior performance can be attributed to several factors. First, the physics-based feature engineering process explicitly incorporates domain knowledge about glucose-specific optical signatures, wavelength-dependent interference patterns, and individual physiological variations. This targeted approach proves more effective than relying on neural networks to automatically discover these relationships from noisy data. Second, the linear regression framework with appropriate regularization provides robustness against overfitting, which becomes particularly important when training data contains noise-stressed levels. Third, the computational efficiency enables real-time implementation on resource-constrained embedded platforms.

The performance degradation observed in neural network approaches under noise-stressed conditions highlights the challenges of applying complex models to noisy, limited datasets. The Optimized PINN, despite incorporating advanced architectural features, shows worse performance than the simpler Original PINN. This suggests that increased model complexity can lead to overfitting when the signal-to-noise ratio is low and training data is limited.

The RTE-based models demonstrate that incorporating more complete physics does not necessarily improve performance under noise-stressed conditions. While the RTE-inspired PINN provides the most physics-heavy description of light transport in this study, its performance is limited by the quality of the underlying data and the challenges of estimating tissue optical properties from limited measurements. The computational cost of RTE modeling may not be justified for practical glucose monitoring applications.

\subsection{Implications for Prototype Development}

The findings of this study have significant implications for practical prototype development. The Enhanced Beer–Lambert model's combination of high accuracy and computational efficiency makes it an ideal candidate for embedded implementation. With only 56 parameters and 0.01 ms inference time, the model can be easily deployed on microcontroller platforms such as the ESP32-S3 while maintaining real-time performance.

The physics-based feature engineering approach provides interpretability and robustness. Unlike black-box neural network approaches, the Enhanced Beer–Lambert model's predictions can be traced back to specific optical principles and physiological parameters, facilitating understanding and validation of the measurement process.

The noise-stressed data generation framework developed in this study provides a valuable tool for prototype development and validation. By modeling hardware limitations, environmental variations, and physiological differences, the framework enables performance assessment before expensive hardware prototyping. This approach can reduce development time and cost while improving the likelihood of successful prototype implementation.

\subsection{Limitations and Future Work}

While this study provides valuable insights into computational approaches for glucose monitoring, several limitations should be acknowledged. The analysis is based on synthetic data, albeit noise-stressed, and is not medical-grade; validation with field measurements will be necessary to confirm the findings. The glucose-NIR correlation of 0.211 achieved in the synthetic data represents our best estimate of low-SNR conditions, but actual prototype implementations may encounter additional challenges not captured in the simulation.

The study focuses on four specific NIR wavelengths (850, 940, 1050, and 1150 nm) selected for optimal glucose sensitivity. Future work should explore the impact of wavelength selection and the potential benefits of additional spectral channels. The hardware model assumes specific component characteristics that may vary across different implementations, and the robustness of the approaches to hardware variations should be investigated.

Future research directions include validation of the Enhanced Beer–Lambert model with field data from prototype implementations, investigation of adaptive calibration techniques to account for individual physiological variations, exploration of hybrid approaches that combine physics-based feature engineering with selective neural network components, and development of uncertainty quantification methods to provide confidence intervals for glucose predictions.

\section{Conclusions}

This comparative analysis of six computational approaches for non-invasive glucose monitoring under noise-stressed synthetic conditions provides several conclusions that inform practical prototype development.

The Enhanced Beer–Lambert model with physics-based feature engineering achieves superior performance compared to all neural network approaches, demonstrating 13.6 mg/dL RMSE while requiring only 56 parameters and 0.01 ms inference time. This represents a 61\% improvement in RMSE compared to the SDNN baseline and establishes a strong benchmark for computational efficiency in glucose monitoring applications.

The noise-stressed synthetic data generation framework successfully models challenging conditions of practical prototype implementation, achieving glucose-NIR correlations of 0.211 that reflect the combined impact of hardware limitations, environmental variations, and physiological individual differences. This framework provides a valuable tool for performance assessment and prototype development.

Physics-based feature engineering proves more effective than complex neural network architectures under noise-stressed conditions, suggesting that domain knowledge incorporation is crucial for robust performance in challenging measurement scenarios. The linear regression framework with appropriate regularization provides sufficient modeling capacity while maintaining robustness against overfitting.

The computational efficiency of the Enhanced Beer–Lambert model makes it suitable for embedded implementation on resource-constrained platforms. The interpretability of the physics-based approach supports transparent analysis of what drives predictions.

These findings provide a practical path forward for prototype development. The Enhanced Beer–Lambert model balances accuracy, efficiency, and interpretability under low-SNR conditions, showing that reliable estimation is possible even when the signal is weak and noise dominates.

\appendix
\section{Model Calibration and Diagnostic Plots}

\begin{figure}[H]
  \centering
  \includegraphics[width=\linewidth]{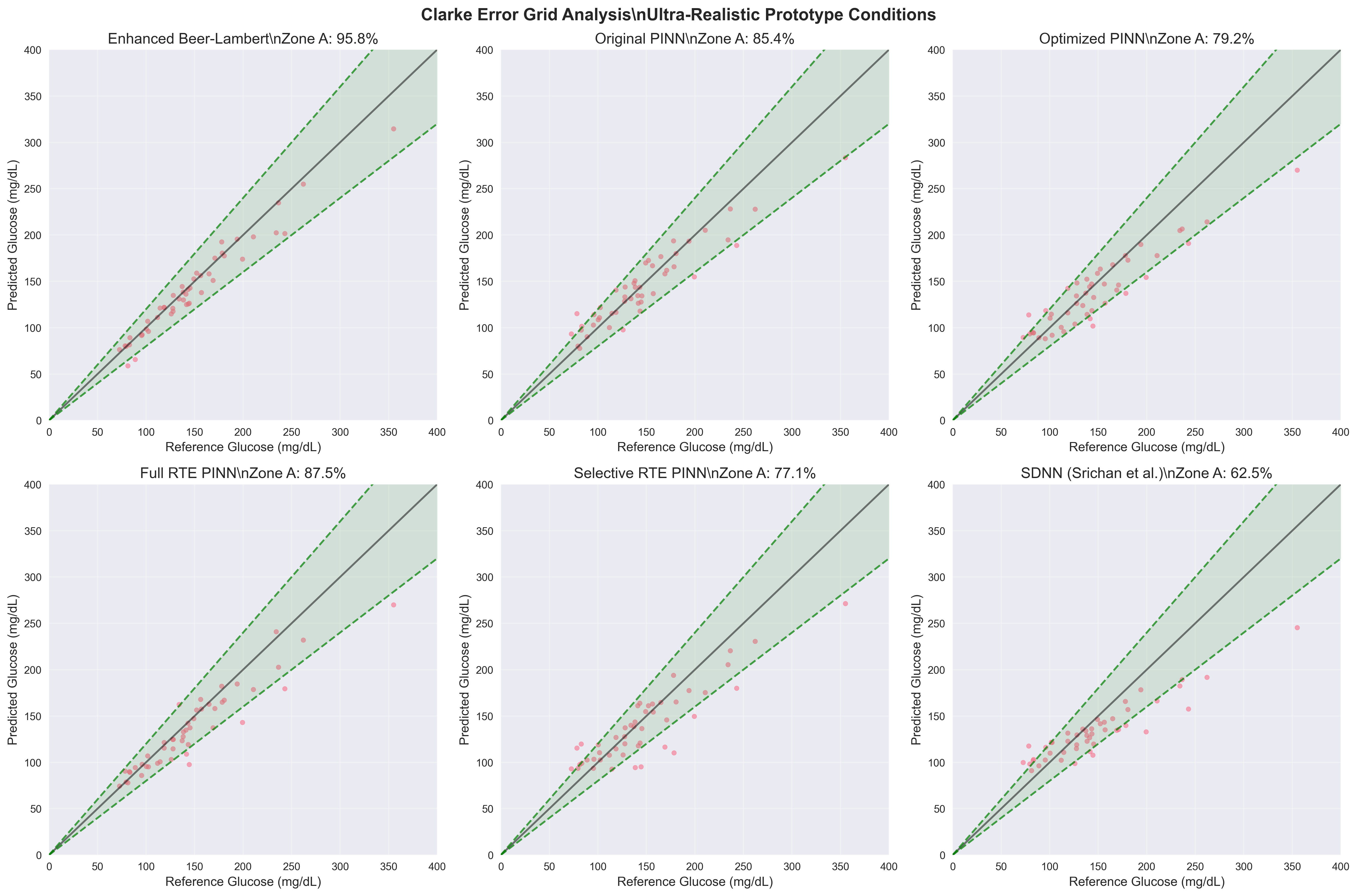}
\caption{Clarke Error Grid (evaluation): the Enhanced Beer–Lambert model shows tighter agreement than neural baselines under noise-stressed conditions.}
  \label{fig:clarke_results}
\end{figure}

\begin{figure}[H]
  \centering
  \includegraphics[width=\linewidth]{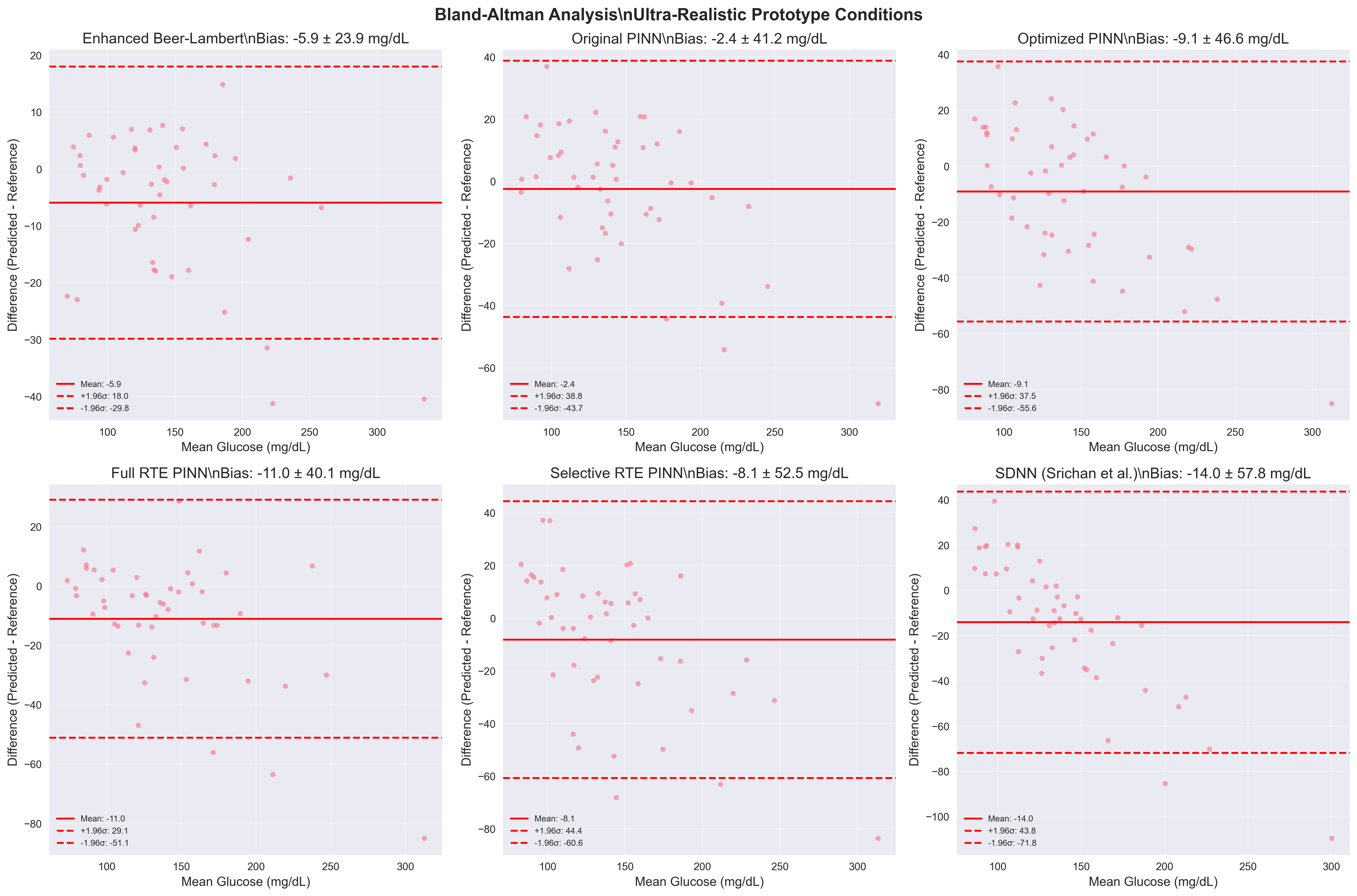}
  \caption{Bland–Altman plots showing bias and limits of agreement across all models. The Enhanced Beer–Lambert model demonstrates tight agreement with minimal bias, while neural models exhibit greater dispersion and systematic error under noise.}
  \label{fig:bland_altman}
\end{figure}

\begin{figure}[H]
  \centering
  \includegraphics[width=\linewidth]{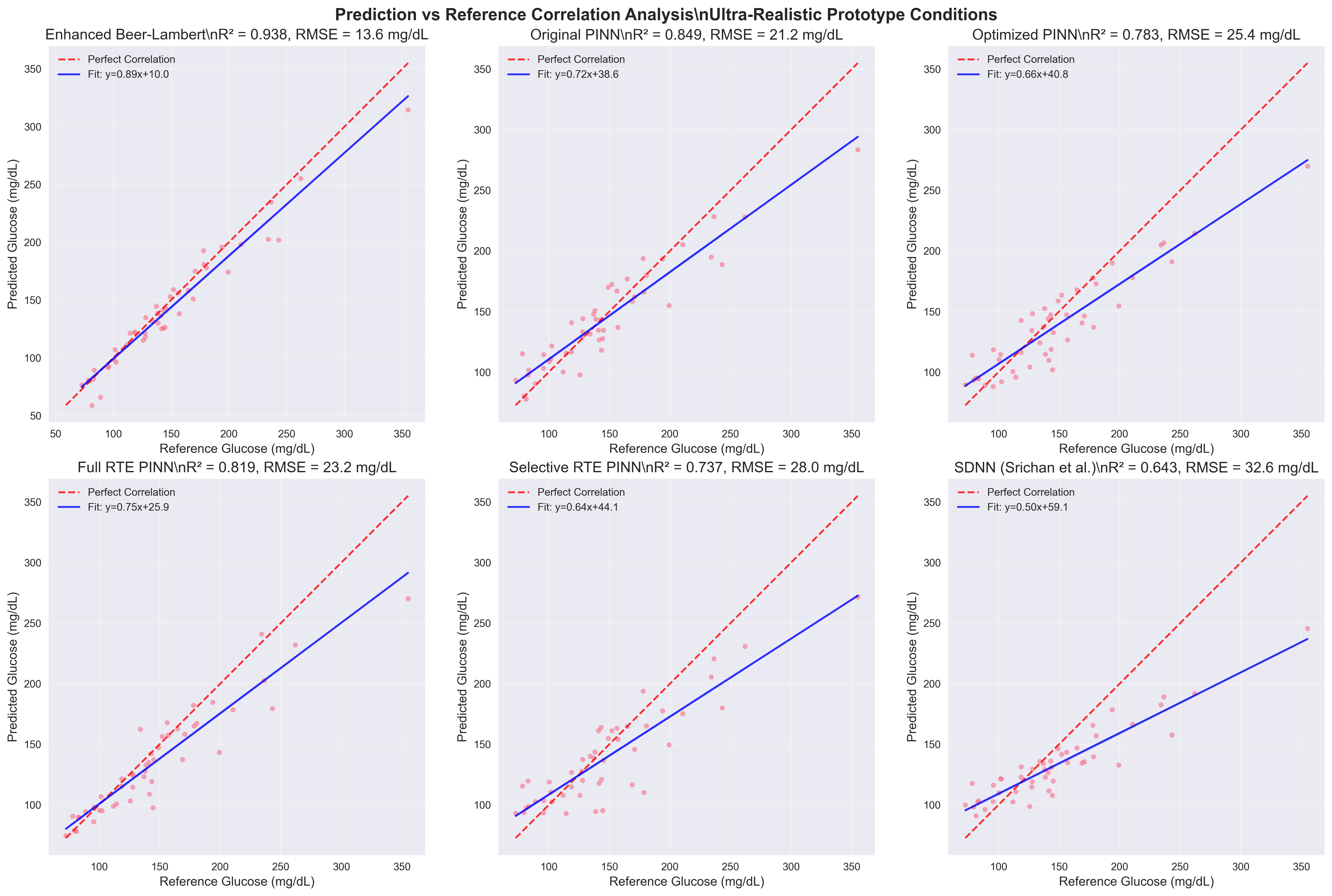}
  \caption{Linearity analysis comparing predicted and ground-truth glucose. The red dashed line shows the ideal unity-slope reference; deviation of the blue regression line from this diagonal visualises systematic calibration bias. Only the Enhanced Beer–Lambert model maintains high linearity across the full range, while deep models suffer compression or offset drift.}
  \label{fig:correlation_linearity}
\end{figure}

\begin{figure}[H]
  \centering
  \includegraphics[width=\linewidth]{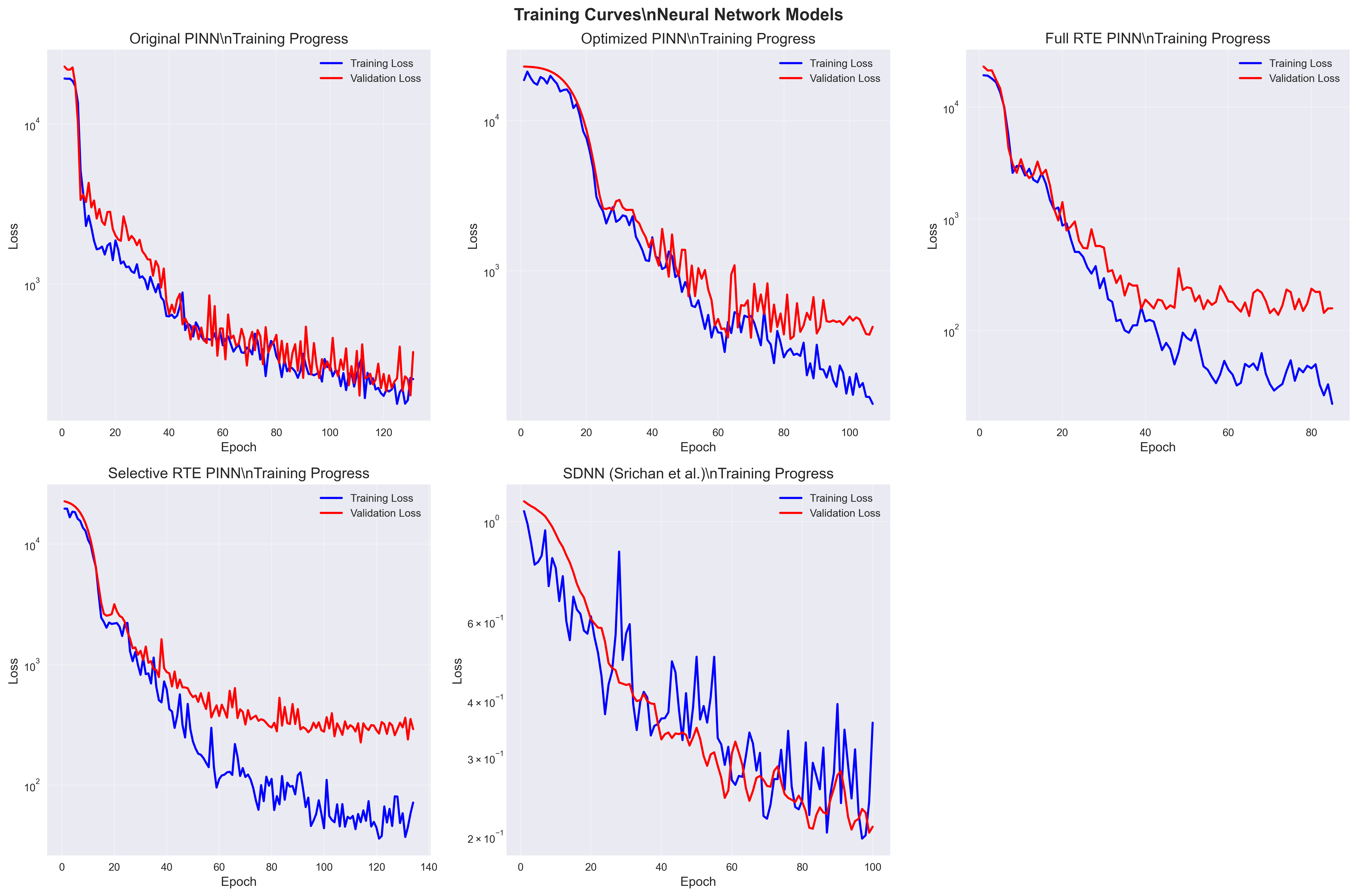}
  \caption{Training dynamics of neural network models. Each subplot shows training and validation loss over epochs, highlighting convergence stability and overfitting risk. The Enhanced Beer–Lambert model is excluded from this plot as it does not require iterative training--its parameters are computed analytically via linear regression on physics-informed features.}
  \label{fig:training_diagnostics}
\end{figure}


\begin{thebibliography}{20}

\bibitem{srichan2022} 
Srichan, C., et al. (2022). Non-invasively accuracy enhanced blood glucose sensor using shallow dense neural networks with NIR monitoring and medical features. \textit{Scientific Reports}, 12, 1769.

\bibitem{arnold1996}
Arnold, M. A., \& Small, G. W. (1996). Non-invasive glucose monitoring. \textit{Current Opinion in Biotechnology}, 7(1), 46-49.

\bibitem{raissi2019}
Raissi, M., Perdikaris, P., \& Karniadakis, G. E. (2019). Physics-informed neural networks: A deep learning framework for solving forward and inverse problems involving nonlinear partial differential equations. \textit{Journal of Computational Physics}, 378, 686-707.

\bibitem{karniadakis2021}
Karniadakis, G. E., et al. (2021). Physics-informed machine learning. \textit{Nature Reviews Physics}, 3(6), 422-440.

\bibitem{beer1852}
Beer, A. (1852). Bestimmung der Absorption des rothen Lichts in farbigen Flüssigkeiten. \textit{Annalen der Physik}, 162(4), 78-88.

\bibitem{amerov2005}
Amerov, A. K., et al. (2005). Scattering and absorption effects in the determination of glucose in whole blood by near-infrared spectroscopy. \textit{Analytical Chemistry}, 77(14), 4587-4594.

\bibitem{tuchin2007}
Tuchin, V. V. (2007). \textit{Tissue optics: light scattering methods and instruments for medical diagnosis}. SPIE Press.

\bibitem{maier2018}
Maier, J. S., Walker, S. A., Fantini, S., Franceschini, M. A., \& Gratton, E. (1994). Possible correlation between blood glucose concentration and the reduced scattering coefficient of tissues in the near infrared.
\textit{Optics Letters}, 19(24), 2062--2064.


\bibitem{khalil2003}
Khalil, O. S. (2003). Non-invasive glucose measurement technologies: an update from 1999 to the dawn of the new millennium. \textit{Diabetes Technology \& Therapeutics}, 6(5), 660-697.

\bibitem{cuomo2022}
Cuomo, S., et al. (2022). Scientific machine learning through physics–informed neural networks: Where we are and what's next. \textit{Journal of Scientific Computing}, 92, Article 88.

\bibitem{lu2021}
Lu, L., Meng, X., Mao, Z., \& Karniadakis, G. E. (2021). DeepXDE: A deep learning library for solving differential equations. \textit{SIAM Review}, 63(1), 208-228.

\bibitem{wang2021}
Wang, S., Teng, Y., \& Perdikaris, P. (2021). Understanding and mitigating gradient flow pathologies in physics-informed neural networks. \textit{SIAM Journal on Scientific Computing}, 43(5), A3055-A3081.

\bibitem{ishimaru1978}
Ishimaru, A. (1978). \textit{Wave propagation and scattering in random media}. Academic Press.

\bibitem{wang1995}
Wang, L., Jacques, S. L., \& Zheng, L. (1995). MCML--Monte Carlo modeling of light transport in multi-layered tissues. \textit{Computer Methods and Programs in Biomedicine}, 47(2), 131-146.

\bibitem{arnold2001}
Arnold, M. A., \& Small, G. W. (1990).
Determination of physiological levels of glucose in an aqueous matrix with digitally filtered Fourier transform near-infrared spectra. \textit{Analytical Chemistry}, 62(14), 1457--1464.

\bibitem{burmeister2000}
Burmeister, J. J., Arnold, M. A., \& Small, G. W. (2000). Noninvasive blood glucose measurements by near-infrared transmission spectroscopy across human tongues. \textit{Diabetes Technology \& Therapeutics}, 2(1), 5-16.

\end{thebibliography}
\end{document}